\documentclass[12pt]{article}
\usepackage{amssymb,amsmath,epsfig}
\begin{document} \parindent=0pt
\parskip=6pt \rm

\begin{center}
{\bf \large Phenomenological study of spin-triplet ferromagnetic\\
superconductors}

{\bf Diana V. Shopova, Tsvetomir E. Tsvetkov, and Dimo I. Uzunov},

 CP Laboratory, Institute of Solid State Physics,\\
Bulgarian Academy of Sciences, BG-1784 Sofia, Bulgaria.
\end{center}

\begin{abstract}
Unconventional superconductivity with spin-triplet Cooper pairing
is reviewed on the basis of the quasi-phenomenological
Ginzburg-Landau theory. The superconductivity, in particular, the
mixed phase of coexistence of ferromagnetism and unconventional
superconductivity is triggered by the spontaneous magnetization.
The mixed phase is stable whereas the other superconducting phases
that usually exist in unconventional superconductors are either
unstable, or, for particular values of the parameters of the
theory, some of these phases are metastable at relatively low
temperatures in a quite narrow domain of the phase diagram. The
phase transitions from the normal phase to the phase of
coexistence is of first order while the phase transition from the
ferromagnetic phase to the coexistence phase can be either of
first or second order depending on the concrete substance. The
Cooper pair and crystal anisotropy are relevant to a more precise
outline  of the phase diagram shape and  reduce the degeneration
of the ground states of the system but they do not drastically
influence the phase stability domains and the thermodynamic
properties of the respective phases.
\end{abstract}

{\bf  Keywords:} superconductivity, ferromagnetism, phase diagram,
order parameter profile, anisotropy. {\bf PACS:} 74.20.De,
74.20.Rp.

\section{Inroduction}

The formation of Cooper pairs with a nonzero angular momentum was
theoretically predicted~\cite{Pitaevskii:1959} in 1959 as a
mechanism of superfluidity of Fermi liquids. In 1972 the same
phenomenon - unconventional superfluidity due to a $p$-wave (spin
triplet) Cooper pairing of $^3$He atoms, was experimentally
discovered in the mK range of temperatures; for details and
theoretical description, see
Refs.~\cite{Leggett:1975,Vollhardt:1990,Volovik:2003}. In contrast
to the standard $s$-wave pairing in usual (conventional)
superconductors, where the electron pairs are formed by an
attractive electron-electron interaction due to a virtual phonon
exchange, the widely accepted mechanism of the Cooper pairing in
superfluid $^3$He is based on an attractive interaction between
the fermions ($^3$He atoms) due to a virtual exchange of spin
fluctuations. Certain spin fluctuation mechanisms of
unconventional Cooper pairing of electrons were proposed also for
the depiction of discovered in 1979 heavy fermion superconductors
(see, e.g., Refs.~\cite{Stewart:1984,Sigrist:1991,Mineev:1999}) as
well as for some classes of high-temperature superconductors (see,
e.g.,
 Refs.~\cite{Sigrist:1987,Annett:1988,Volovik:1988,Blagoeva:1990,Uzunov:1990,
 Uzunov:1993,Annett:1995,Harlingen:1995,Tsuei:2000}).

The possible superconducting phases in unconventional superconductors
are described in the framework of the general Ginzburg-Landau (GL)
effective free energy functional~\cite{Uzunov:1993} with the help of
the  symmetry groups theory. A variety of possible superconducting
orderings were predicted for different crystal
structures~\cite{Volovik0:1985,Volovik:1985,Ueda:1985,Blount:1985,Ozaki:1985,Ozaki:1986}.
A detailed thermodynamic analysis~\cite{Blagoeva:1990,Volovik:1985} of
the homogeneous (Meissner) phases and a renormalization group
investigation~\cite{Blagoeva:1990} of the superconducting phase
transition up to the two-loop approximation have been also performed
(for a three-loop renormalization group analysis, see
Ref.~\cite{Antonenko:1994}; for effects of magnetic fluctuations and
disorder, see~\cite{Busiello:1991,Busiello:1990}).

In 2000, experiments~\cite{Saxena:2000} at low temperatures ($T \sim 1$
K) and high pressure ($T\sim 1$ GPa) demonstrated the existence of spin
triplet superconducting states in the metallic compound UGe$_2$. This
superconductivity is triggered by the spontaneous magnetization of the
ferromagnetic phase which exists at much higher temperatures and
coexists with the superconducting phase in the whole domain of
existence of the latter below $T \sim 1$ K; see also experiments
published in Refs.~\cite{Huxley:2001,Tateiwa:2001}, and  the discussion
in Ref.~\cite{Coleman:2000}. Moreover, the same phenomenon of existence
of superconductivity at low temperatures and high pressure in the
domain of the $(T,P)$ phase diagram where the ferromagnetic order is
present has been observed in other ferromagnetic metallic compounds
(ZrZn$_2$~\cite{Pfleiderer:2001} and  URhGe~\cite{Aoki:2001}) soon
after the discovery~\cite{Saxena:2000} of superconductivity in UGe$_2$.

In contrast to other superconducting materials, for example,
ternary and Chevrel phase compounds, where the effects of magnetic
order on superconductivity are also substantial (see,
e.g.,~\cite{Vonsovsky:1982,Maple:1982,Sinha:1984,Kotani:1984}), in
these ferromagnetic compounds the phase transition temperature
($T_f$) to the ferromagnetic state is much higher than the phase
transition temperature ($T_{FS})$ from ferromagnetic to a (mixed)
state of coexistence of ferromagnetism and superconductivity. For
example, in UGe$_2$ $T_{FS}$ is $0.8$ K whereas the critical
temperature of the phase transition from paramagnetic to
ferromagnetic state in the same material is $T_f =35
$K~\cite{Saxena:2000,Huxley:2001}. It can be supposed that in such
substances the material parameter $T_s$ defined as the (usual)
critical temperature of the second order phase transition from
normal to uniform (Meissner) supercondicting state in a zero
external magnetic field is much lower than the phase transition
temperature $T_{FS}$. Note, that the mentioned
 experiments with the compounds~UGe$_{2}$, URhGe, and ZrZn$_2$
do not give any evidence for the existence of a standard
normal-to-superconducting phase transition in a zero external
magnetic field.

Moreover, it seems that the superconductivity in the metallic
compounds, mentioned above, always coexists with the ferromagnetic
order and is enhanced by it. As claimed in Ref.~\cite{Saxena:2000}
in these systems the superconductivity seems to arise from the
same electrons that create the band magnetism, and is most
naturally understood as a triplet rather than spin-singlet pairing
phenomenon. Note, that all three metallic compounds, mentioned so
far, are itinerant ferromagnets.  A similar type of unconventional
superconductivity has been suggested~\cite{Saxena:2001} as a
possible outcome of recent experiments in Fe~\cite{Shimizu:2001},
where a superconducting phase is discovered at temperatures below
$2$ K for pressures between 15 and 30 GPa. Note, that both vortex
and Meissner superconductivity phases~\cite{Shimizu:2001} have
been found in the high-pressure crystal modification of Fe which
has a hexagonal close-packed crystal structure. In this hexagonal
lattice the strong ferromagnetism of the usual bcc iron crystal
probably disappears~\cite{Saxena:2001}. Thus one can hardly claim
that there is a coexistence of ferromagnetism and
superconductivity in Fe but the clear evidence of a
superconductivity is also a remarkable achievement.

The important point in all discussions of the interplay between
superconductivity and ferromagnetism is that a small amount of
magnetic impurities can destroy superconductivity in conventional
($s$-wave) superconductors by breaking up the ($s$-wave) electron
pairs with opposite spins (paramagnetic impurity
effect~\cite{Abrikosov:1960}). In this aspect the phenomenological
arguments~\cite{Ginzburg:1956} and the conclusions on the basis of
the microscopic theory of magnetic impurities in $s$-wave
superconductors~\cite{Abrikosov:1960} are in a complete agreement
with each other; see, e.g.,
Refs.~\cite{Vonsovsky:1982,Maple:1982,Sinha:1984,Kotani:1984}. In
fact, a total suppression of conventional ($s$-wave)
superconductivity should occur in the presence of an uniform
spontaneous magnetization $\mbox{\boldmath$M$}$, i.e., in a
standard ferromagnetic phase~\cite{Ginzburg:1956}. The physical
reason for this suppression is the same as in the case of magnetic
impurities, namely, the opposite electron spins in the $s$-wave
Cooper pair turn along the vector $\mbox{\boldmath$M$}$ in order
to lower their Zeeman energy and, hence, the pairs break down.
Therefore, the ferromagnetic order can hardly coexist with
conventional superconducting states. Especially this is so for the
coexistence of uniform superconducting and ferromagnetic states
when the superconducting order parameter
$\psi(\mbox{\boldmath$x$})$ and the magnetization
$\mbox{\boldmath$M$}(\mbox{\boldmath$x$})$ do not depend on the
spatial vector $\mbox{\boldmath$x$}$.

But yet a coexistence of $s$-wave superconductivity and
ferromagnetism may appear in uncommon materials and under quite
special circumstances. Furthermore, let us emphasize that the
conditions for the coexistence of nonuniform (``vertex'',
``spiral'', ``spin-sinosoidal'' or ``helical'') superconducting
and ferromagnetic states are less restrictive than those for the
coexistence of uniform superconducting and ferromagnetic orders.
Coexistence of nonuniform phases has been discussed in details,
 in particular, experiment and theory of ternary and Chevrel-phase
compounds, where such coexistence seems quite likely; for a
comprehensive review, see, for example, Refs.
~\cite{Vonsovsky:1982,Maple:1982,Sinha:1984,Kotani:1984,Buzdin:1983}.

In fact, the only two superconducting systems for which the
experimental data allow assumptions in favor of a coexistence of
superconductivity and ferromagnetism are the rare earth ternary
boride compound ErRh$_4$B$_4$ and the Chervel phase compound
HoMo$_6$S$_8$; for a more extended review, see
Refs.~\cite{Maple:1982,Machida:1984}. In these compounds the phase
of coexistence most likely appears in a very narrow temperature
region just below the Curie temperature $T_f$ of the ferromagnetic
phase transition. At lower temperatures the magnetic moments of
the rare earth 4$f$ electrons become better aligned, the
magnetization increases and the $s$-wave superconductivity pairs
formed by the conduction electrons disintegrate.

We shall not extend our consideration over all important aspects
of the long standing problem of coexistence of superconductivity
and ferromagnetism  rather we shall concentrate our attention on
the description of the newly discovered coexistence of
ferromagnetism and unconventional (spin-triplet) superconductivity
in the itinerant ferromagnets UGe$_2$, ZrZn$_2$, and URhGe. Here
we wish to emphasize that the main object of our discussion is the
superconductivity of these compounds and, in a second place in
rate of importance we put the problem of coexistence. The reason
is that the existence of superconductivity in such itinerant
ferromagnets is a highly nontrivial phenomenon. As noted in
Ref.~\cite {Machida:2001} the superconductivity in these materials
seems difficult to explain in terms of previous theories~\cite
{Vonsovsky:1982,Maple:1982,Kotani:1984} and requires new concepts
for the interpretation of experimental data.

We have already mentioned that in ternary compounds the
ferromagnetism comes from the localized 4$f$ electrons whereas the
s-wave Cooper pairs are formed by conduction electrons. In UGe$_2$
and URhGe the 5$f$ electrons of U atoms form both superconducting
and ferromagnetic order~\cite{Saxena:2000,Aoki:2001}. In ZrZn$_2$
the same
 twofold role is played by the 4$d$ electrons of Zr.
Therefore the task is to describe this behavior of the band electrons
at a microscopic level. One may speculate about a spin-fluctuation
mediated unconventional Cooper pairing as is in case of $^3$He and
heavy fermion superconductors. These important issues have not yet a
reliable answer and for this reason we shall confine our consideration
to a phenomenological level.

Reliable experimental data, for example, the data about the
coherence length and the superconducting
gap~\cite{Saxena:2000,Huxley:2001,Aoki:2001,Pfleiderer:2001}, are
in favor of the conclusion about a spin-triplet Cooper pairing in
these metallic compounds, although the mechanism of the pairing
remains unclear. We shall essentially use this reliable
conclusion. Besides, this point of view is consistent with the
experimental observation of coexistence of superconductivity only
in the low temperature part of the ferromagnetic domain of the
phase diagram ($T,P$)which means that a pure (non ferromagnetic)
superconducting phase has not been observed. This circumstance is
also in favor of the assumption of a spin-triplet
superconductivity.  Our investigation leads to results which
confirm this general picture.

On the basis of the experimental data and conclusions presented
for the first time in Refs.~\cite{Saxena:2000,Coleman:2000} and
shortly afterwards confirmed in
Refs.~\cite{Huxley:2001,Tateiwa:2001,Pfleiderer:2001,Aoki:2001} it
can be accepted that the  superconductivity in these magnetic
compounds is considerably enhanced by the ferromagnetic order
parameter $\mbox{\boldmath$M$}$ and, perhaps, it could not exist
without this ``mechanism of ferromagnetic trigger,'' or, in short,
``$\mbox{\boldmath$M$}$-trigger.'' Such phenomenon is possible for
spin-triplet Cooper pairs, where the electron spins point parallel
to each other and their turn along the vector of the spontaneous
magnetization $\mbox{\boldmath$M$}$ does not cause a
 break down of the spin-triplet Cooper pairs but rather stabilizes them and,
perhaps, stimulates their creation. We shall describe this phenomenon
at a phenomenological level.

Recently, the  phenomenological theory that explains the
coexistence of ferromagnetism and unconventional spin-triplet
superconductivity of GL type was developed
~\cite{Machida:2001,Walker:2002}. The possible low-order couplings
between the superconducting and ferromagnetic order parameters
were derived with the help of general symmetry group arguments and
several important features of the superconducting vortex state in
the ferromagnetic phase of unconventional ferromagnetic
superconductors were established~\cite{Machida:2001,Walker:2002}.

In this paper we shall use the approach presented in
Refs.~\cite{Machida:2001,Walker:2002}  to investigate the
conditions for the occurrence of the Meissner phase and to
demonstrate that the presence of ferromagnetic order enhances the
$p$-wave superconductivity. Our consideration is focused on the
ground state, namely, we are interested in uniform phases, where
the order parameters (the superconducting order parameter $\psi$
and the magnetization vector $M = \left\{M_j, j = 1,2,3\right\}$),
do not depend on the spatial vector $\vec{x}\in V $, where $V$ is
the volume of the system. Recent results about the phase diagram
and the phase transitions~\cite{Shopova1:2003,Shopova2:2003}, and
thermodynamic quantities~\cite{Shopova3:2003} will be essentially
used in our investigation.

Our study is based on the mean-field
approximation~\cite{Uzunov:1993} as well as on familiar results
for the possible phases in nonmagnetic superconductors with
triplet ($p$-wave) Cooper
pairs~\cite{Blagoeva:1990,Uzunov:1990,Volovik:1985}. Results from
Refs.\cite{Shopova1:2003,Shopova2:2003,Shopova3:2003,Shopova:2004}
will be reviewed and extended. In our preceding investigation
\cite{Shopova1:2003,Shopova2:2003,Shopova3:2003} both Cooper pair
anisotropy and crystal anisotropy have been neglected in order to
clarify the main effect of the coupling between the ferromagnetic
and superconducting order parameters. The phenomenological GL free
energy is quite complex and the inclusion of these anisotropies is
related with lengthy formulae and a multivariant analysis which
obscures the final results. Here we shall point our attention to
the effect of the Cooper pairs anisotropy.

There exists a formal similarity between the phase diagram we have
obtained  and the phase diagram of certain improper
ferroelectrics~\cite{Gufan:1980,Gufan:1981,Latush:1985,Toledano:1987,Gufan:1987,Cowley:1980}.
The variants of the theory of improper ferroelectrics, known
before 1980, were criticized in Ref.~\cite{Cowley:1980} for their
oversimplification and inconsistency with the experimental
results. But the further development of the theory has no such
disadvantages (see, e.g., Ref.~\cite{Toledano:1987,Gufan:1987}).
We should emphasize that the symmetry of the GL model of
spin-triplet ferromagnetic superconductors is quite different from
the symmetry of known models in ferroelectrics and, hence, the
results for ferroelectric systems can hardly be applied to
superconductors without additional investigations.

\section{Ginzburg-Landau free energy}

Consider the GL free energy $F(\psi,\mbox{\boldmath$M$})=V f(\psi,
\mbox{\boldmath$M$})\:$, where the free energy density
$f(\psi,\mbox{\boldmath$M$})$ (for short hereafter called ``free
energy'') of a spin-triplet ferromagnetic superconductor is given by

\begin{eqnarray}
\label{eq2} f(\psi, \mbox{\boldmath$M$}) &= &
 a_s|\psi|^2 +\frac{b_s}{2}|\psi|^4 + \frac{u_s}{2}|\psi^2|^2 +
\frac{v_s}{2}\sum_{j=1}^{3}|\psi_j|^4 +
a_f\mbox{\boldmath$M$}^2 + \frac{b_f}{2}\mbox{\boldmath$M$}^4\\
\nonumber && + i\gamma_0 \mbox{\boldmath$M$}.(\psi\times \psi^*) +
\delta \mbox{\boldmath$M$}^2 |\psi|^2\;.
\end{eqnarray}

In Eq.~(1), $\psi = \left\{\psi_j;j=1,2,3\right\}$ is a
three-dimensional complex vector ($\psi_j = \psi_j^{\prime} + i
\psi_j^{\prime\prime}$) describing the unconventional
(spin-triplet) superconducting order and $\mbox{\boldmath$B$} =
(\mbox{\boldmath$H$} + 4\pi\mbox{\boldmath$M$}) = \nabla \times
\mbox{\boldmath$A$}$ is the magnetic induction;
$\mbox{\boldmath$H$} = \{H_j; j = 1,2,3\}$ is the external
magnetic field, $\mbox{\boldmath$A$} = \left\{A_j;
j=1,2,3\right\}$ is the magnetic vector potential ($\nabla.\:A =
0$). In Eq.~(1), $b_s > 0$, $b_f > 0$, $a_f = \alpha_f(T-T_f)$ is
given by the positive material parameter $\alpha_f$ and the
ferromagnetic critical temperature $T_f$ corresponding to a simple
ferromagnet ($\psi \equiv 0$), and $a_s = \alpha_s(T-T_s)$, where
$\alpha_s$ is another positive material parameter and $T_s$ is the
critical temperature of a standard second order phase transition
which may occur at $|H| = {\cal{M}} = 0$; ${\cal{M}} =
|\mbox{\boldmath$M$}|$. The parameter $u_s$ describes the
anisotropy of the spin-triplet Cooper pair whereas the crystal
anisotropy is described by the parameter
$v_s$~\cite{Blagoeva:1990,Volovik:1985}.

The two orders -- the magnetization vector $M = \left\{M_j\right\}$ and
$\psi = \left\{A_j\right\}$, interact through the last two terms in
(1). The $\gamma_0-$term~\cite{Walker:2002} ensures the triggering of
the superconductivity by the ferromagnetic order ($\gamma_0>0$) whereas
the $\delta-$term makes the model more realistic in the strong coupling
limit~\cite{Machida:2001}. Both $\psi M$-interaction terms included in
(1) are important for a correct description of the temperature-pressure
($T,P$) phase diagram of the ferromagnetic
superconductor~\cite{Shopova1:2003,Shopova2:2003}. In general, the
parameter $\delta$ for ferromagnetic superconductors may take both
positive and negative values.

As we are interested in the ground state properties, we set the
external magnetic field equal to zero $(H = 0)$. Besides, we emphasize
that the magnetization vector $M$ may produce vortex superconducting
phase in case of type II superconductivity. The investigation of
nonuniform (vortex) states can be made with the help of gradient terms
in the free energy which take into account the spatial variations of
the order parameter field $\psi$. This task is beyond our present
consideration. Rather we investigate the basic problem about the
possible stable uniform (Meissner) superconducting phases which may
coexist with uniform ferromagnetic order. For this aim the free energy
(1) is quite convenient.

In case of a strong easy axis type of magnetic anisotropy, as is
in UGe$_2$~\cite{Saxena:2000}, the overall complexity of
mean-field analysis of the free energy $f(\psi,
\mbox{\boldmath$M$})$ can be avoided by performing an
``Ising-like'' description: $\mbox{\boldmath$M$} =
(0,0,{\cal{M}})$. Further, because of the equivalence of the
``up'' and ``down'' physical states $(\pm \mbox{\boldmath$M$})$
the thermodynamic analysis can be done within the ``gauge"
${\cal{M}} \geq 0$. But this stage of consideration can also be
achieved without the help of crystal anisotropy arguments. When
the magnetic order has a continuous symmetry one may take
advantage of the symmetry of  the total free energy $f(\psi,
\mbox{\boldmath$M$})$ and avoid the study of equivalent
thermodynamic states that occur as a result of the respective
symmetry breaking at the phase transition point but they have no
effect on thermodynamics of the system. In the isotropic system
one may again choose a gauge, in which the magnetization vector
has the same direction as  $z$-axis ($|\mbox{\boldmath$M$}| = M_z
= {\cal{M}}$) and this will not influence the generality of
thermodynamic analysis. Here we shall prefer the alternative
description within which the ferromagnetic state may occur as two
equivalent ``up'' and ``down'' domains with magnetizations $
{\cal{M}}$ and $ -{\cal{M}}$, respectively.

We shall use adequate notations which reduce the number of
 parameters. With the help of
\begin{equation}
\label{eq10} b = (b_s + u_s + v_s) > 0
\end{equation}
we redefine the order parameters and the other parameters in the
following way:
\begin{eqnarray}
\label{eq11} &&\varphi_j =b^{1/4}\psi_j = \phi_je^{\theta_j}\:,\;\;\; M
= b_f^{1/4}{\cal{M}}\:,\\ \nonumber && r =
\frac{a_s}{\sqrt{b}}\:,\;\;\; t =\frac{a_f}{\sqrt{b_f}}\:, \;\;\; w =
\frac{u_s}{b}\:, \;\;\; v =\frac{v_s}{b}\:, \\ \nonumber &&\gamma=
\frac{\gamma_0}{b^{1/2}b_f^{1/4}}\:,\;\;\; \gamma_1=
\frac{\delta}{(bb_f)^{1/2}}\:.
\end{eqnarray}

Having in mind our approximation of uniform $\psi$ and
$\mbox{\boldmath$M$}$
 and the notations~(2)~-~(3), the free energy density
 $f(\psi,M)$ can be written in the form
\begin{eqnarray}
\label{eq12} f(\psi,M)& = & r\phi^2 + \frac{1}{2}\phi^4
  + 2\gamma\phi_1\phi_2 M \mbox{sin}(\theta_2-\theta_1) + \gamma_1 \phi^2 M^2
+ tM^2 + \frac{1}{2}M^4\\ \nonumber && -2w
\left[\phi_1^2\phi_2^2\mbox{sin}^2(\theta_2-\theta_1)
 +\phi_1^2\phi_3^2\mbox{sin}^2(\theta_1-\theta_3) +
 \phi_2^2\phi_3^2\mbox{sin}^2(\theta_2-\theta_3)\right] \\ \nonumber
&& -v[\phi_1^2\phi_2^2 + \phi_1^2\phi_3^2 + \phi_2^2\phi_3^2].
\end{eqnarray}
In the above free energy the order parameters $\psi$ and
$\mbox{\boldmath$M$}$ are defined per unit volume.

 We assume that $T_f > T_s$. This is
the case when the superconductivity is triggered by the magnetic
order. We shall discuss the stable phases in the temperature
region $T
> T_s$. The case $T_f < T_s$ may be also important for neutron
stars so it needs a special investigation. When $T_s \sim T_f$  a
quite simple analytical treatment is possible. All these cases may
be of interest to the description of ferromagnetic
superconductivity in stellar objects whereas in condensed matter
only case of $T_f \gg T_s$ has been observed so far.

We work in the framework of the standard mean-field
analysis~\cite{Uzunov:1993}. The stable phases correspond to
global minima of the GL energy (1). The equilibrium phase
transition line separating two phases is defined by the
thermodynamic states, where the respective GL free energies are
equal.

\section{Phases}

We shall not enumerate and discuss all phases described by
Eq.~(1). Rather we shall focus our attention on the stable phases
at relatively high temperatures $(T > T_s)$. The calculations show
that for temperatures $T> T_s$, i.e., for $r> 0$, there are three
stable phases. Two of them are quite simple: the normal phase
($\psi = M = 0$) with existence and stability domains given by
$t>0$ and $r>0$, and the ferromagnetic phase (FM) given by $\psi
=0$ and $M^2 = -t$ whose  existence condition is $ t<0$,the is
stability domain defined by the inequalities $r
> \gamma_1t$ and
\begin{equation}
\label{eq39} r > \gamma_1t + \gamma\sqrt{-t}\:.
\end{equation}
The third stable phase is the phase of coexistence of
superconductivity and ferromagnetism (hereafter referred to as
FS). This phase is the main object of our consideration. It is
given by the following equations:
\begin{equation}
\label{eq40} \phi_1 = \phi_2=\frac{\phi}{\sqrt{2}}\:, \;\;\; \phi_3 =
0\:,
\end{equation}
\begin{equation}
\label{eq41} \phi^2= (\pm \gamma M-r-\gamma_1 M^2)\:,
\end{equation}
\begin{equation}
\label{eq42} (1-\gamma_1^2)M^3\pm \frac{3}{2} \gamma \gamma_1 M^2
+\left(t-\frac{\gamma^2}{2}-\gamma_1 r\right)M \pm \frac{\gamma
r}{2}=0\:,
\end{equation}
and
\begin{equation}
\label{eq43} (\theta_2 - \theta_1) = \mp \frac{\pi}{2} + 2\pi k\:,
\end{equation}
($k = 0, \pm 1,...$). The upper sign in Eqs.~(7)~--~(9)
corresponds to a domain where $\mbox{sin}(\theta_2-\theta_1) = -1$
and the lower sign corresponds to a second domain which may be
referred to as FS$^{*}$; there $\mbox{sin}(\theta_2-\theta_1) =
1$. These two domains are equivalent and describe the same
ordering. We shall focus on the upper sign in (7)~--~(9), i.e. on
FS. The phase diagram ($t,r$) has qualitatively the same shape as
the phase transition lines corresponding to $w = 0$ but there are
essential quantitative differences between them. We shall discuss
them in the next section. Note, that the system exhibits both
first and second order phase transitions and complex phase
transition points: triple and tricritical
points~\cite{Shopova1:2003,Shopova:2004}.

\section{Anisotropy effects}

Our analysis demonstrates that when the anisotropy of the Cooper
pairs is taken into account, there will be not drastic changes in
the shape the phase diagram for $r>0$ and the order of the
respective phase transitions. Of course, there will be some
changes in the size of the phase domains and the formulae for the
thermodynamic quantities. Besides, and this seems to be the main
anisotropy effect, the $w$- and $v$-terms in the free energy lead
to a stabilization of the order along the main crystal directions
which, in other words, means that the degeneration of the possible
ground states is considerably reduced. So there will be a smaller
number of marginally stable states.

Let us neglect the crystal anisotropy by setting $v_s=0$ in
Eq.~(1) and concentrate our attention on the Cooper pair
anisotropy described by the $u_s$-term in the GL model. The
dimensionless anisotropy parameter $w \sim u_s$ given by Eq.~(3)
can be either positive or negative depending on the sign of $u_s$.
Obviously when $ u_s > 0$, the parameter $w$ will be positive too
($0< w<1$). We shall illustrate the influence of Cooper-pair
anisotropy in this case. The order parameters ($M$, $\phi_j$,
$\theta_j$) are given by Eqs.~(6), (9),
 \begin{equation}
\label{eq54} \phi^2=\frac{\pm \gamma M-r-\gamma_1 M^2}{(1-w)} \ge 0\:,
\end{equation}
and
\begin{equation}
\label{eq55} (1- w - \gamma_1^2)M^3 \pm \frac{3}{2} \gamma \gamma_1 M^2
+\left[t(1-w)-\frac{\gamma^2}{2}-\gamma_1 r\right]M \pm \frac{\gamma
r}{2}=0\:,
\end{equation}
where the meaning of the upper and lower sign is the same as explained
just below Eq.~(9). We consider the FS domain corresponding to the
upper sign in the Eq.~(10) and (11). The stability conditions for FS
read,
\begin{equation}
\label{eq56} \frac{ (2-w)\gamma M- r -\gamma_1M^2}{1-w} \ge 0\:,
\end{equation}
\begin{equation}
\label{eq57} \gamma M -wr-w \gamma_1 M^2 \ge 0\:,
\end{equation}
and
\begin{equation}
\label{eq58} \frac{1}{1-w}\left[3(1-w-\gamma_1^2) M^2 + 3 \gamma
\gamma_1 M + t(1-w)-\frac{\gamma^2}{2} -\gamma_1 r \right]\geq 0\:.
\end{equation}
For $M\ne (\gamma/2 \gamma_1)$ we can express the function
$r(M,t)$ defined by Eq.~(11), substitute the obtained expression
for $r(M,t)$ in the existence (10) and stability conditions
(12)-(14) and do the analysis in the same way as for
$w=0$~\cite{Shopova:2004}. The most substantial qualitative
difference between the cases $w>0$ and $w<0$ is that for $w < 0$
the stability of FS is bounded for $r<0$.

\section{Conclusion}

We have done an investigation of the M-trigger effect in
unconventional ferromagnetic superconductors. This effect due to
the $M\psi_1\psi_2$-coupling term in the GL free energy consists
of bringing into existence of superconductivity in a domain of the
phase diagram of the system that is entirely in the region of
existence of the ferromagnetic phase. This form of coexistence of
unconventional superconductivity and ferromagnetic order is
possible for temperatures above and below the critical temperature
$T_s$, which corresponds to the standard phase transition of
second order from normal to Meissner phase -- usual uniform
superconductivity in a zero external magnetic field, which appears
outside the domain of existence of the ferromagnetic order. Our
investigation has been mainly intended to clarify the
thermodynamic behaviour at temperatures $T_s< T < T_f$, where the
superconductivity cannot appear without the mechanism of
M-triggering. We have described the possible ordered phases (FM
and FS) in this most interesting temperature interval.

The Cooper pair and crystal anisotropies have also been
investigated and their main effects on the thermodynamics of the
triggered phase of coexistence are established. In discussions of
concrete real materials one should take into account the
respective crystal symmetry but the variation of the essential
thermodynamic properties with the change of the type of symmetry
is not substantial when the low symmetry and low order (in both
$M$ and $\psi$) $\gamma$-term is present in the free energy.

Below the superconducting critical temperature $T_s$ a variety of pure
superconducting and mixed phases of coexistence of superconductivity
and ferromagnetism exists and the thermodynamic behavior at these
relatively low temperatures is more complex than in known cases of
improper ferroelectrics. The case $T_f < T_s$ also needs a special
investigation. Our results are referred to the possible uniform
superconducting and ferromagnetic states. Vortex and other nonuniform
phases need a separate study.

The relation of the present investigation to properties of real
ferromagnetic compounds, such as UGe$_2$, URhGe, and ZrZn$_2$, has
been discussed throughout the text. In these real compounds the
ferromagnetic critical temperature is much larger than the
superconducting critical temperature $(T_f \gg T_s)$ and that is
why the M-triggering of the spin-triplet superconductivity is very
strong. Moreover, the $\gamma_1$-term is important to stabilize
the FM order up to the absolute zero (0 K), as is in the known
spin-triplet ferromagnetic superconductors. The
neglecting~\cite{Walker:2002} of the symmetry conserving
$\gamma_1$-term hinders the proper description of real substances
of this type. More experimental information about the values of
the material parameters ($a_s, a_f, ...$) is required in order to
outline the thermodynamic behavior and the phase diagram in terms
of thermodynamic parameters $T$ and $P$. In particular, a reliable
knowledge about the dependence of the parameters $a_s$ and $a_f$
on the pressure $P$, the value of the characteristic temperature
$T_s$ and the ratio $a_s/a_f$ at zero temperature are of primary
interest.

{\bf Acknowledgments:}

The authors warmly greet Reinhard Folk with his 60th anniversary !
D.V.S. and D.I.U. thank Reinhard Folk for the valuable
collaboration, discussions, and professional support over the last
fifteen years.

\end{document}